\documentclass[fleqn,usenatbib]{mnras}

\usepackage{graphicx}
\usepackage{graphics}
\usepackage{amsmath, amssymb}
\usepackage{multirow}
\usepackage{color}
\usepackage{xcolor}
\usepackage{bm}		
\usepackage[normalem]{ulem}
\usepackage{comment}
\usepackage{gensymb}
\usepackage{soul}
\usepackage{flushend}
\usepackage{lastpage}

\usepackage{float}

\def\kms{{\rm\,km\,s^{-1}}}

\def\kpc{{\rm\,kpc}}
\def\pc{{\rm\,pc}}

\def\mathnew{\mathsurround=0pt}   
\def\simov#1#2{\lower .5pt\vbox{\baselineskip0pt  
    \lineskip-.5pt\ialign{$\mathnew#1\hfil##\hfil$\crcr#2\crcr\sim\crcr}}}

\def\'#1{\ifx#1i{\accent"13\i}\else{\accent"13#1}\fi}

\def\aap{{A\&A}}

\def\apj{{ApJ}}
\def\apjl{{ApJL}}
\def\nat{{Nature}}
\def\mnras{{MNRAS}}
\def\apjs{{ApJS}}

\defcitealias{Poggio2021}{P21}
\defcitealias{Poggio2022}{P22}

\title[3D metal-rich arms \& Extended Radcliffe Wave]{Metal-rich stellar counterpart of the Radcliffe Wave and the 3D chemical footprints of the Milky Way spiral arms}

\author[L. Martinez-Medina et al.]{Luis Martinez-Medina$^{1}$\thanks{Contact e-mail:\href{mailto:lamartinez@astro.unam.mx}{lamartinez@astro.unam.mx}}\href{https://orcid.org/0000-0002-5749-8255}{\includegraphics[scale=0.5]{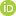}},
Eloisa Poggio$^{2}$\href{https://orcid.org/0000-0003-3793-8505}{\includegraphics[scale=0.5]{orcid.png}},
Elizabeth Moreno-Hilario$^{3}$\href{https://orcid.org/0000-0002-6906-2379}{\includegraphics[scale=0.5]{orcid.png}} \\
$^{1}$Instituto de Astronom\'ia, Universidad Nacional Aut\'onoma de M\'exico, A.P. 70-264, 04510. Ciudad de México, México\\
$^{2}$INAF - Osservatorio Astrofisico di Torino, via Osservatorio 20, 10025 Pino Torinese (TO), Italy\\
$^{3}$Department of Astronomy, Tsinghua University, Beijing 100084, People’s Republic of China\\
}
\date{Released \today}

\pubyear{2025}
\begin{document}
\label{firstpage}
\pagerange{\pageref{firstpage}--\pageref{lastpage}}
\maketitle

\begin{abstract}

Mapping the Milky Way spiral arms in the vertical direction remains a challenging task that has received little attention. Taking advantage of recent results that link the position of the Galactic spiral arms to metal-rich regions in the disc, we analyse a sample of young giant stars from {\it Gaia} DR3 and use their metallicity distribution to produce a 3D metallicity excess map. The map shows signatures of the spiral arms, whose vertical height vary across the Galactic disc, reaching up to 400$\pc$ in amplitude and exhibiting vertical asymmetries with respect to the mid-plane. Specifically, the Perseus arm displays a high vertical asymmetry consistent with the Galactic warp.
Moreover, we find evidence of a metal-rich stellar structure that undulates vertically, nearly in phase with the arrangement of star-forming regions named the Radcliffe Wave. This new structure is larger and extends beyond the Radcliffe Wave, reaching vertical amplitudes of $\sim$ 270$\pc$ and extending for at least 4$\kpc$ in length. We confirm that for at least half of its length this Extended Radcliffe Wave is the inner edge of the Local Arm. The finding of a metal-rich stellar counterpart of the Radcliffe Wave shows that mapping the three-dimensional metallicity distribution of young stellar populations reveals key information about the structures and chemical enrichment in the Galactic disc.

\end{abstract}                

 
\section{Introduction} 
\label{sec:intro}

Galactic cartography has reached a level of detail and precision that has not been seen before. With every data release, the European Space Agency {\it Gaia} mission \citep[][]{GaiaCollaboration:2016, GaiaCollaboration:2018, GaiaCollaboration:2021, GaiaCollaboration:2023} has shown clearer evidence of a perturbed stellar disc \citep[][]{Antoja2018,Hunt2024}, imprints of major mergers in the Galaxy \citep[][]{Belokurov2018,Laporte2019,Vasiliev2021}, chemical inhomogeneities that in turn trace the different Galactic components \citep[][]{Recio-Blanco2023,Lian2024}, and disequilibrium features in the distribution and kinematics of stars \citep[see][for a recent review]{Hunt_Vasiliev2025}.

The mapping of the Milky Way has improved significantly in one particular area, the revealing of the three-dimensional (3D) morphology of the disc and the non-axisymmetric structures in it. For example, it has been possible to obtain a better characterization of the warp in the outer disc \citep[e.g.][]{Lemasle2022,Dehnen2023,Cabrera-Gadea2024}, the discovery of vertical undulations and corrugations across the disc \citep[e.g.][]{Alves2020,Poggio2024,Barnes2025}, or to get a detailed picture of the spiral arms as traced by a variety of objects \citep[][]{Eilers2020,Castro-Ginard2021,Poggio2021,Poggio2022,Martinez-Medina2022,Palicio2023}. However, a 3D map of the stellar disc is far from complete, and the {\it Gaia} high-resolution data (in combination with other catalogues) contains an important part of this information yet to be revealed.

Of particular interest is the study of the 3D morphology of the spiral arms, a property that has received little attention due to the lack of reliable tracers capable of defining and following the vertical boundary of these non-axisymmetric structures. In this context, the recent results from \citet[][hereafter \citetalias{Poggio2021}]{Poggio2021} and \citet[][hereafter \citetalias{Poggio2022}]{Poggio2022} are of special interest. \citetalias{Poggio2021} utilized astrometry and photometry from {\it Gaia} EDR3 to trace the spatial distribution of young stellar populations, revealing coherent spiral arm segments extending several kiloparsecs across the Galactic disc. Expanding on this, \citetalias{Poggio2022} mapped the chemical inhomogeneities of three different samples of giant stars in the Galactic disc and found that the youngest population exhibits large scale metallicity variations that strongly correlate with the location of the spiral arms detected as overdensities in \citetalias{Poggio2021}. Such azimuthal variations in the metallicity distribution have been confirmed by \citet{Hawkins2023} using a combination of stars from LAMOST and {\it Gaia}, and by \citet{Hackshaw2024} using giant stars from APOGEE. More recently \citet{Barbillon2025} not only found that stars within spiral arms are more metal-rich but also more [Ca/Fe]-poor and [MG/Fe]-poor than inter-arm stars.
These results open the possibility of using the chemical footprint of the spiral arms as tracer of their morphology not only on the mid-plane of the Galaxy but also along the vertical direction.

In disc galaxies, spiral arms are predicted to induce chemical azimuthal variations, as shown by numerical simulations \citep{Grand:2016,Khoperskov2018, Khoperskov2022, Khoperskov2023, Debattista2025, Graf2025} and analytic chemical evolution models \citep{Spitoni2019,Spitoni2023}. However, the exact correspondence between the current distribution of chemical elements and the spiral structure in the Galactic disc is unknown: while there is a tendency of spiral arms to be more metal-rich than the inter-arm regions, local deviations from a one-to-one correlation can exist \citep[see for example Fig. 3 of][]{Khoperskov2018}. To further complicate the picture, chemical azimuthal variations can also be generated by radial migration induced by satellite galaxies \citep{Carr2022} and the central bar \citep{DiMatteo2013, Filion2023}. For this reason, studying the three-dimensional distribution of stellar metallicity in the Galactic disc can reveal fundamental clues on the structure and chemical evolution of the Milky Way. Specifically, young stars can give us a unique perspective on the star formation and enrichment history of the spiral structure in the Milky Way disc. 

In this work we study the 3D metallicity distribution of young giant stars in the solar neighbourhood, and focus on creating a 3D map of the spiral arm segments that can be mapped in a complementary way by these chemical inhomogeneities. Then, pushing forward this approach, we revisit the recently discovered Radcliffe Wave to show its connection to a larger vertically undulating metal-rich stellar structure. 

This paper is organized as follows: in Section \ref{sec:data} we describe the {\it Gaia} DR3 stellar sample. In Section \ref{sec:3dmaps} we present 3D maps of the Local, Perseus, Sagittarius-Carina and Scutum spiral arms. In Section \ref{sec:RW} we explore the vertical stellar undulation along the extension of the Radcliffe Wave. Finally, in Section \ref{sec:conclusions} we present our conclusions.

\section{Data sample}
\label{sec:data}

In this work we decided to use one of the stellar samples presented in \citetalias{Poggio2022}, specifically Sample A, as it shows the strongest correlation between regions of metallicity excess in the disc and regions of spiral arm overdensities. Here we give a brief summary of the data sample while a detailed description can be found in \citetalias{Poggio2022}.

Sample A in \citetalias{Poggio2022} consist of a subsample of giant stars from {\it Gaia} DR3, selected to filter out those sources with larger uncertainties in spectroscopic and astrometric determinations. The selected stars are brighter than the red clump and the position they occupy in the $T_{\rm eff} - {\rm log}(g)$ diagram places their typical age at 100~Myr or younger. A kinematical cut was applied to the velocities to remove possible halo stars, leaving a sample characterized by velocity dispersions $(\sigma_{V_R}, \sigma_{V_{\phi}}, \sigma_{V_Z}) = (28.6, 26.6, 19.5)\kms$. Regarding the spatial distribution, the majority of stars are located within 4$\kpc$ from the Sun and are constrained to a vertical position of $|Z|<0.75\kpc$. Here we adopt $(R,Z)_{\odot} = (8.249,0.0208)\kpc$ for the Sun's Galactocentric position \citep{GRAVITY2021,Bennett2019}, and $(V_R,V_{\phi},V_Z)_{\odot}=(-9.5,250.7,8.56)\kms$ for the Sun's Galactocentric cylindrical velocities \citep{Reid2020,GRAVITY2021}.   

With this young and dynamically cold sample, \citetalias{Poggio2022} constructed a mean metallicity map $\langle[{\rm M/H}]\rangle$ projected on the $X-Y$ plane. Aside from the expected higher metallicity at inner radii, the authors find evidence of azimuthal variations in the metallicity distribution (see Fig. 1 from \citetalias{Poggio2022}). Then, to get a better contrast of these metallicity substructures \citetalias{Poggio2022} define a new variable, the metallicity excess $\langle[{\rm M/H}]\rangle_{\rm loc} - \langle[{\rm M/H}]\rangle_{\rm large}$, where the local metallicity $\langle[{\rm M/H}]\rangle_{\rm loc}$ is averaged using a smoothing Gaussian kernel over a scale-length of 0.175$\kpc$, and the large scale metallicity $\langle[{\rm M/H}]\rangle_{\rm large}$ is averaged over a scale 7 times larger than the local. The choice of these smoothing scale-lengths is discussed in Appendix \ref{appendix:smoothing}. In Figure \ref{fig:MH_Excess_xy} we reproduce the $X-Y$ metallicity excess map for Sample A, where red (blue) structures represent regions where the metallicity is above (below) the average. In this map three red diagonal-like features stand out, one at the upper left, another one at the central region, and a third one at the bottom right.

\begin{figure}
\begin{center}
\includegraphics[width=\columnwidth]{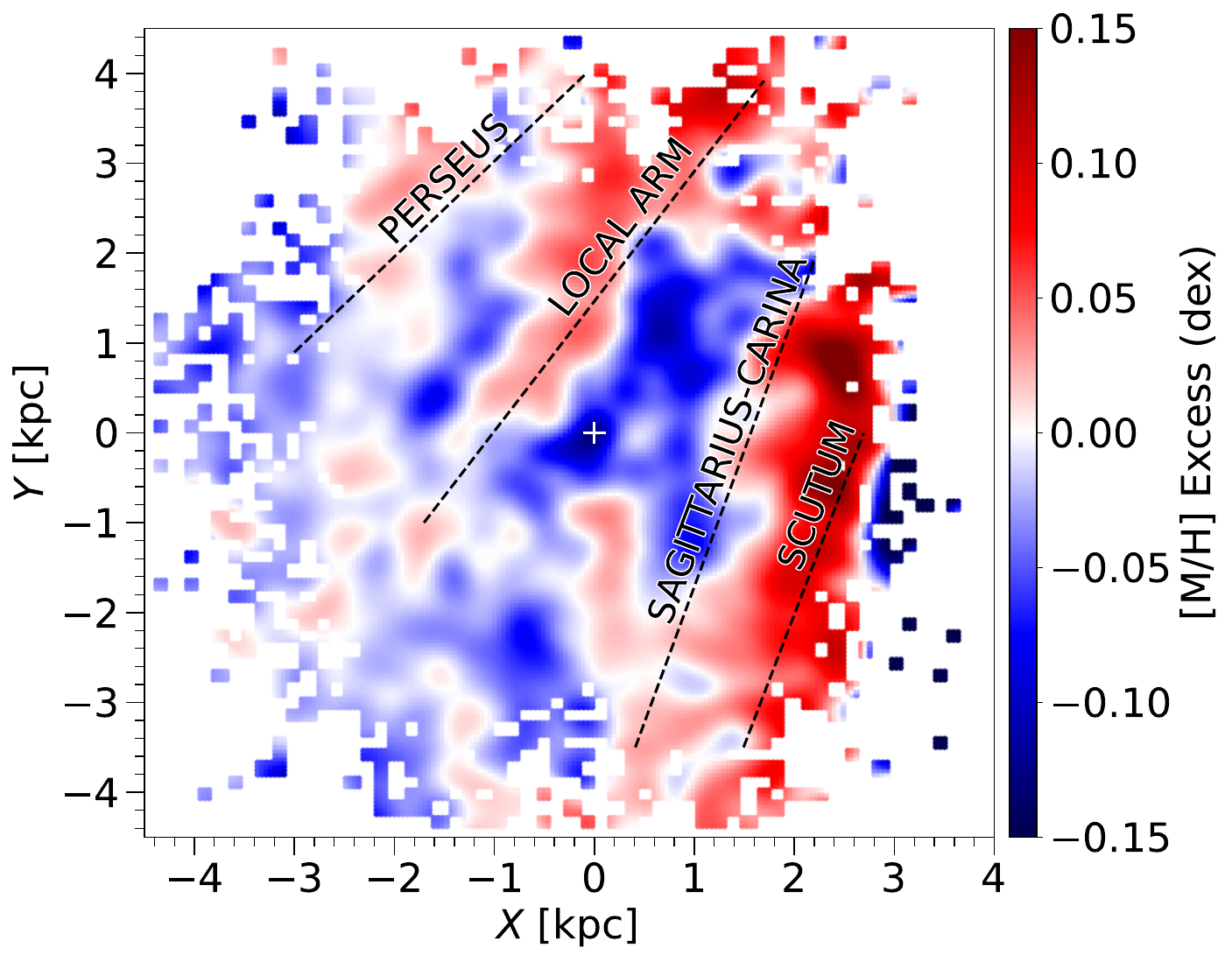}
\end{center}
\caption{Map of [M/H] excess for Sample A. The dashed lines roughly follow the location of arm segments traced by stellar overdensities (\citetalias{Poggio2021}). The white cross marks the Sun's position at (0,0).}
\label{fig:MH_Excess_xy}
\end{figure}

One of the main results of \citetalias{Poggio2022} is that when the map of metallicity excess is compared with the map of spiral arm overdensities computed in \citetalias{Poggio2021} for upper main sequence stars, the location of the red structures in Figure \ref{fig:MH_Excess_xy} show a strong positive correlation with the location of Perseus, Local, and Sagittarius-Carina spiral arm segments. This result not only shows the strong influence of the spiral arms on the chemo-dynamics of the young stellar component, but reveals the stellar metallicity as a potential tracer of the arms location and morphology, as well as the enrichment history of the young Galactic disc.

\section{The spiral arms in 3D}
\label{sec:3dmaps}

Given the strong correlation of the positive metallicity excess of Sample A with the location of the spiral arms, here we use the metallicity excess as a key diagnostic, which brings information about the 3D morphology and chemical enrichment of the spiral arms. 

The first step is to produce a 3D map of metallicity excess following the procedure presented in \citetalias{Poggio2022} and extending it to the vertical dimension. Hence, we compute $\langle[{\rm M/H}]\rangle_{\rm loc} - \langle[{\rm M/H}]\rangle_{\rm large}$ in a 3D grid, where $\langle[{\rm M/H}]\rangle_{\rm loc}$ and $\langle[{\rm M/H}]\rangle_{\rm large}$ are averaged using a smoothing Gaussian kernel for each position $(X, Y, Z)$ in the grid, adopting local scale-lengths of $h_X=h_Y=0.175\kpc$ (as in \citetalias{Poggio2022}) and $h_Z=0.1\kpc$. While for the large scale-lengths the values in each direction are 7 times that of the local ones. Then, we keep only the regions with positive metallicity excess, which tend to be correlated with the morphology of the spiral arms. We note, however, that local deviations between the 3D distribution of spiral arms and positive metallicity excess regions can exist, even in a scenario where the spiral arms are the main drivers of azimuthal metallicity variations in the Galactic disc \citep[see for example Figure 3 in][]{Khoperskov2018}. For this reason, our maps should not be taken as direct tracers of the 3D structure of the spiral arms, but, instead, as a new diagnostic on their 3D chemical impact.

\begin{figure*}
\begin{center}
\includegraphics[width=\textwidth]{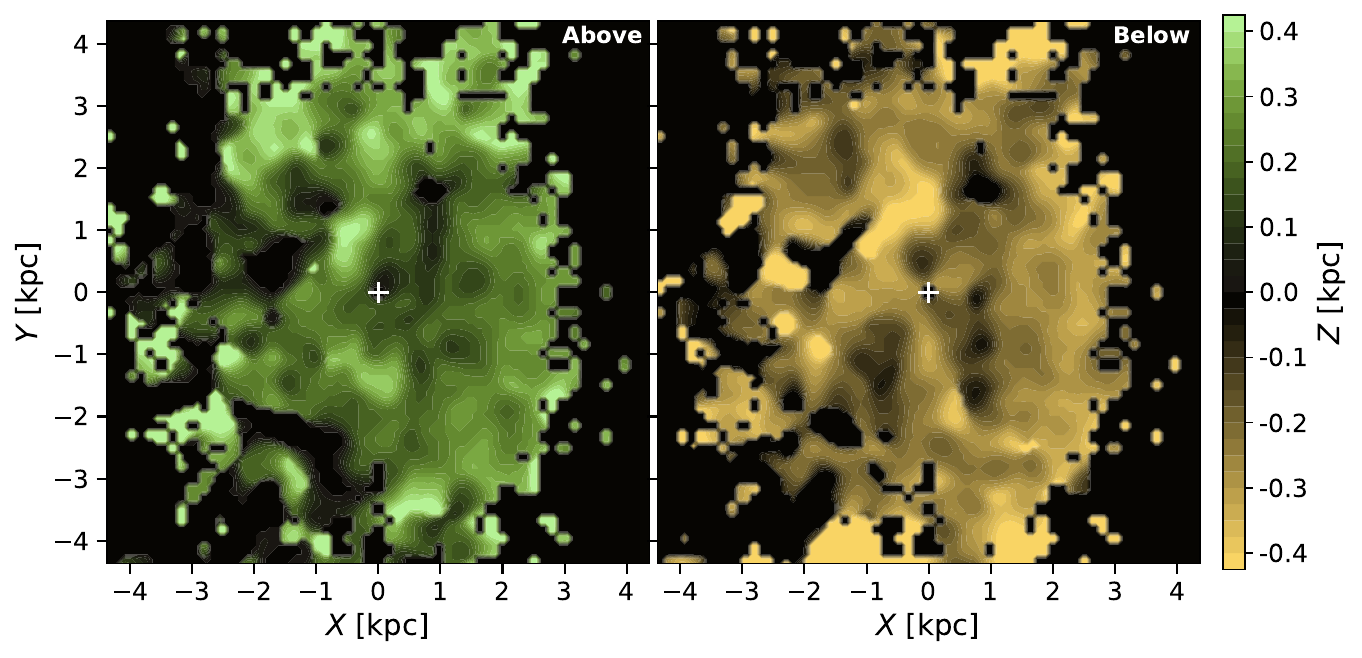}
\end{center}
\caption{ 3D map of the spiral arms as traced by positive stellar metallicity excess. The colour code and contours of equal elevation show {\it relief} features of the spiral arms seen from above the disc (left), and below the disc (right). The spiral arms exhibit a changing vertical thickness along their extension, and look different from above and below, revealing vertically asymmetric arms.    }
\label{fig:Elevation_Hillshade}
\end{figure*}

To show the 3D distribution of the metal-rich stellar regions, in Figure \ref{fig:Elevation_Hillshade} we show the envelope of all positive metallicity excess seen from above the plane (left panel) and below the plane (right panel), where the colour code indicates the maximum vertical separation, $Z$, from the mid-plane. This map already reveals the complex structure of the spiral arms traced by metallicity. They display a changing vertical height along and across their extension, while a comparison between the two panels shows that the arms are not symmetric with respect to the $Z=0$ plane. The overall picture is a series of crests and valleys, where the arm segments associated to Perseus, Local, Sagittarius-Carina and Scutum stand out above and below the mid-plane.

To look into the undulatory nature of the spiral arms with more detail we focus on the arm segments that better correlate with a positive metallicity excess. We proceed by defining four vertical planes that cut through the 3D metallicity excess distribution across the directions illustrated with dashed lines in Figure \ref{fig:MH_Excess_xy}. The choice of these specific directions roughly reproduces the position of the spiral arms according to \citetalias{Poggio2021}, where the arms are traced by stellar overdensities. We provide interactive 3D Figures that better illustrate the vertical cuts and the resulting cross-sections of the spiral arms\footnote[1]{Interactive 3D figures for a more intuitive view of the spiral arms and the Extended Radcliffe Wave \url{https://luismartinez-astro.github.io/Martinez-Medina_2025/Gallery.html}}.

\begin{figure*}
\begin{center}
\includegraphics[width=\textwidth]{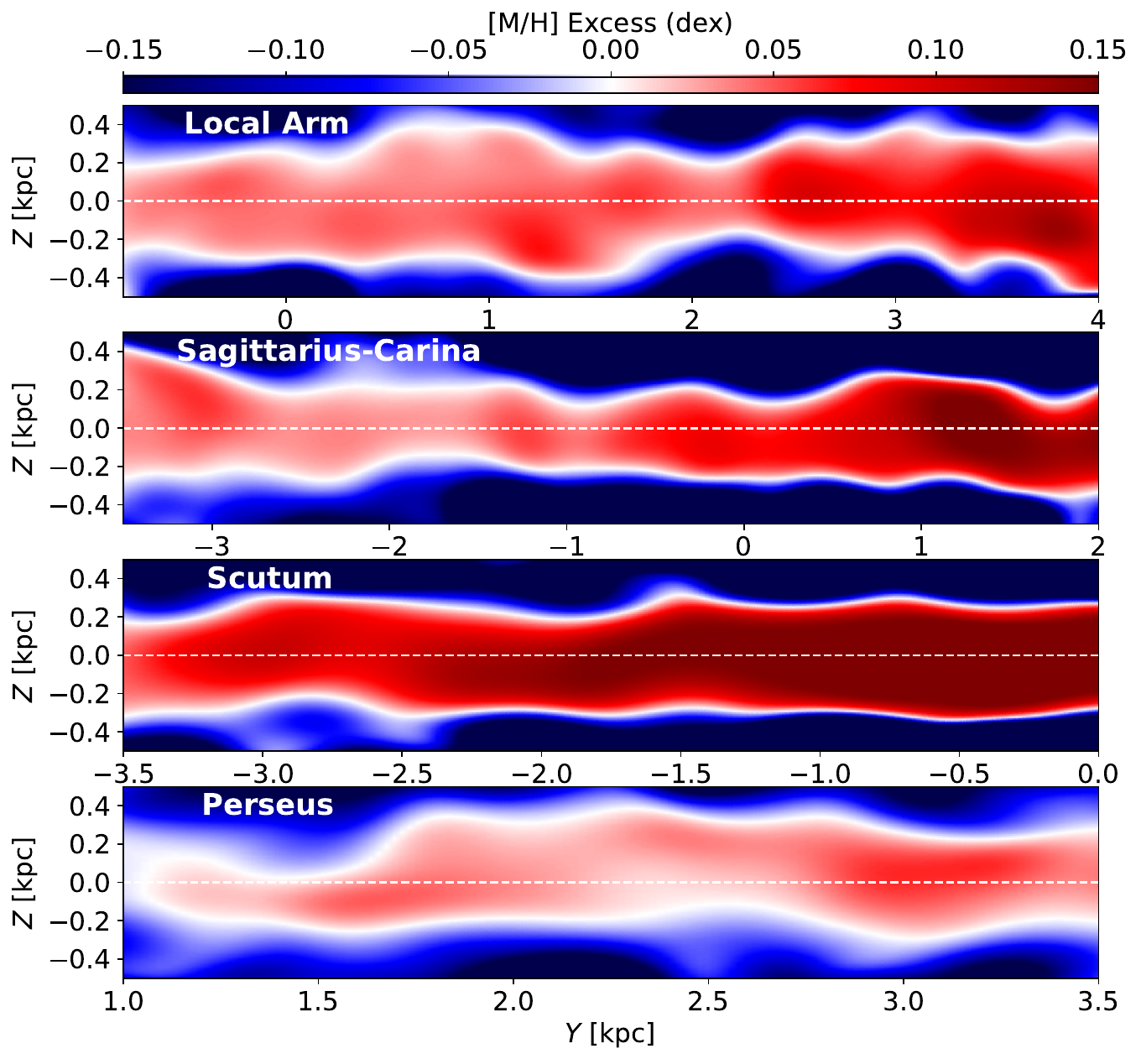}
\end{center}
\caption{Vertical cross-sections of the 3D metallicity excess for the Local, Sagittarius-Carina, Scutum, and Perseus spiral arms, respectively. Note that the range in $Y$ values is different for each arm. }
\label{fig:Arms_yz}
\end{figure*}

As a result, Figure \ref{fig:Arms_yz} shows $Y-Z$ vertical slices of the 3D metallicity excess for the Local, Sagittarius-Carina, Scutum, and Perseus spiral arms, respectively. It is worth mentioning that these slices are not 2D projections, but rather cross-section views of the 3D [M/H] excess when cutting through it. Here, the vertical fluctuations in the morphology of the spiral arms are revealed in more detail. The Local Arm reaches amplitudes of about 400$\pc$ and exhibits several changes in its vertical thickness. Part of the vertical fluctuations and clumpiness might be real, but part might be due to extinction and statistical fluctuations. This behaviour is more pronounced for Sagittarius-Carina, with larger thickness changes and asymmetries. On the other hand, Scutum oscillates with more regularity, and is fairly symmetric along the analysed segment. Finally, the Perseus segment shows a high vertical asymmetry, with most of the metal-rich stars in this arm located above the mid-plane for $Y \gtrsim 1 \kpc$; this is the reason why the Perseus segment hosting the Cassiopeia region (i.e. $X\sim -2\kpc$, $Y\sim 2.5\kpc$) stands out more in the left panel of Figure \ref{fig:Elevation_Hillshade} compared to the right panel. We interpret this difference as due to the Galactic warp: in that region of the disc, the warp of the Milky Way is expected to bend the disc slightly upwards with respect to the $Z=0$ plane, causing a more pronounced disc signature above than below the plane. 

On the other hand, no clear signature of the warp is detected on the other side of the outer disc (i.e. $Y<0$, $X<0$), because the metallicity excess features in our sample at that position are less pronounced than in the rest of the Galactic disc, making a three-dimensional mapping more challenging. Aside from the major vertical variations, it is possible that small-scale undulations are related to statistical fluctuations of the sample. Future works based on more numerous datasets will reveal further details on this point.

Thanks to the all-sky coverage of Gaia GSP-spec metallicites \citep[see Fig. 1 and 2 in][]{Recio-Blanco2023}, no major biases are expected to be present in the maps shown here. However, we note that interstellar extinction can potentially make an impact, especially in the regions towards the Galactic center. Specifically, the decrease in metallicity (and metallicity excess) at the very inner edge of our map is presumably due extinction, due to which we miss the low-latitudes metal-rich stars in the disc, and only see stars at high latitudes, which tend to be relatively poor in metals.

\section{The Extended Radcliffe Wave}
\label{sec:RW}
 
Recently, a vertically oscillating structure known as the Radcliffe Wave, was detected as a narrow coherent arrangement of star-forming clouds \citep{Alves2020}; it is located in the solar neighbourhood, with a wave-like shape of maximum amplitude of about 160$\pc$ and a total extension of $\sim 2.7\kpc$. Its well characterized spatial distribution makes it an structure of interest to validate the method of using the [M/H] excess distribution as tracer of the morphology of dynamically cold stellar structures. Therefore, we now explore the region of the Galactic disc where the Radcliffe Wave is contained and apply the same analysis done for the spiral arms. 

\begin{figure}
\begin{center}
\includegraphics[width=\columnwidth]{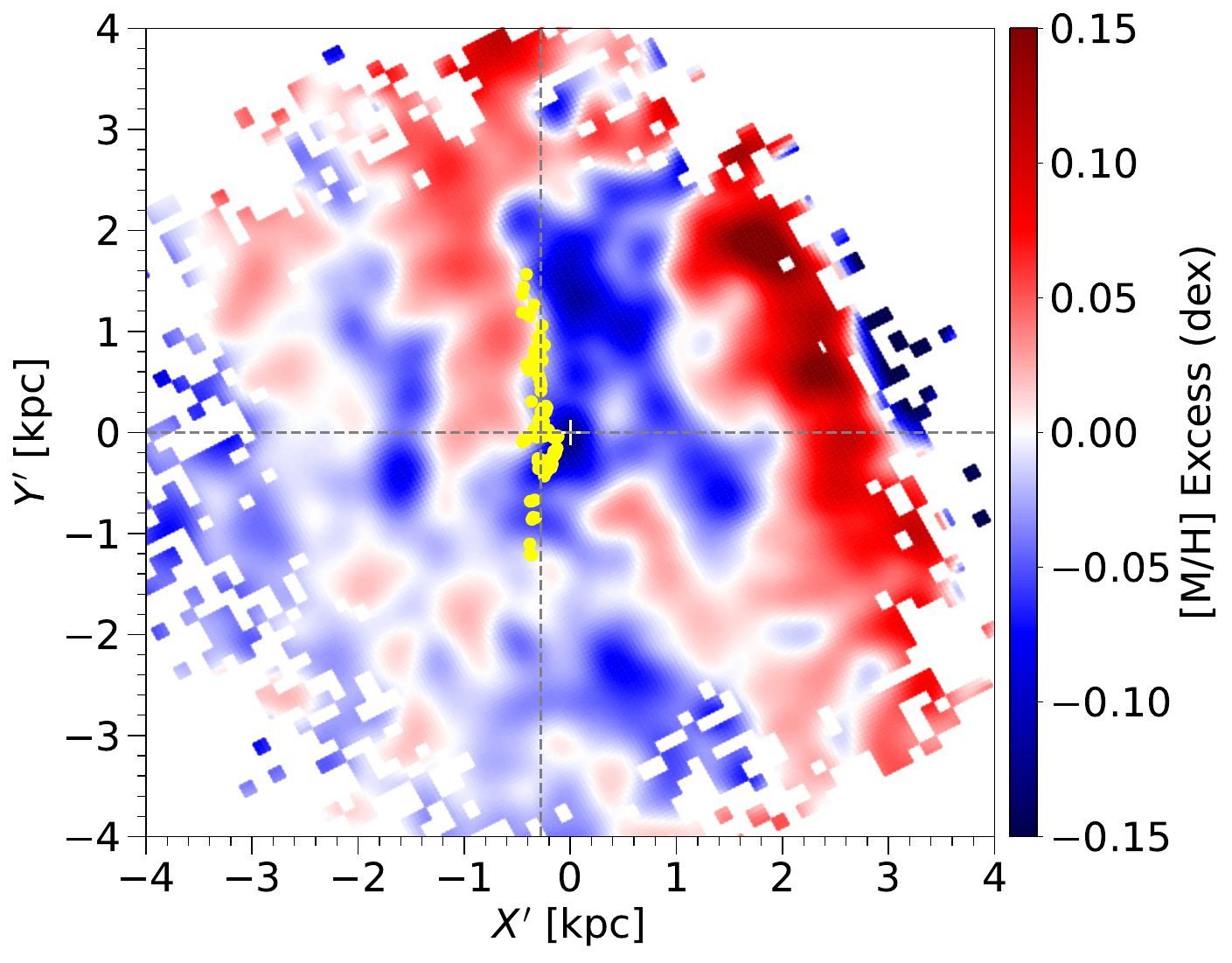}
\end{center}
\caption{[M/H] Excess map from Fig. \ref{fig:MH_Excess_xy} with the original axes rotated anticlockwise by 28$\degree$. The yellow symbols are the star-forming clouds from \citet{Alves2020}, while the grey dashed lines indicate the directions followed to make a longitudinal and a transversal cut of the 3D [M/H] excess.}
\label{fig:MH_Excess_xy_rotated}
\end{figure}

To illustrate the portion of the disc that will be analysed, Figure \ref{fig:MH_Excess_xy_rotated} shows the same metallicity excess map from Figure \ref{fig:MH_Excess_xy}, and as a reference we plot the location of the star-forming clouds from \citet{Alves2020}. Note, however, that this map is rotated anticlockwise by 28$\degree$ in comparison to the original axes, such that in the new $X^{\prime}$ and $Y^{\prime}$ axes the clouds complexes lie mostly along the $Y^{\prime}$ axis \citep[similar to the lower panels of Fig. 2 in ][]{Thulasidharan:2022}. This makes it easier to define a line that passes through the arrange of clouds complexes and allows for a better comparison with the morphology of the Radcliffe Wave. Figure \ref{fig:MH_Excess_xy_rotated} also shows a vertical and a horizontal dashed line, which mark the directions we use to make a longitudinal and transversal cut of the 3D [M/H] excess distribution and produce edge-on views.

\begin{figure*}
\begin{center}
\includegraphics[width=\textwidth]{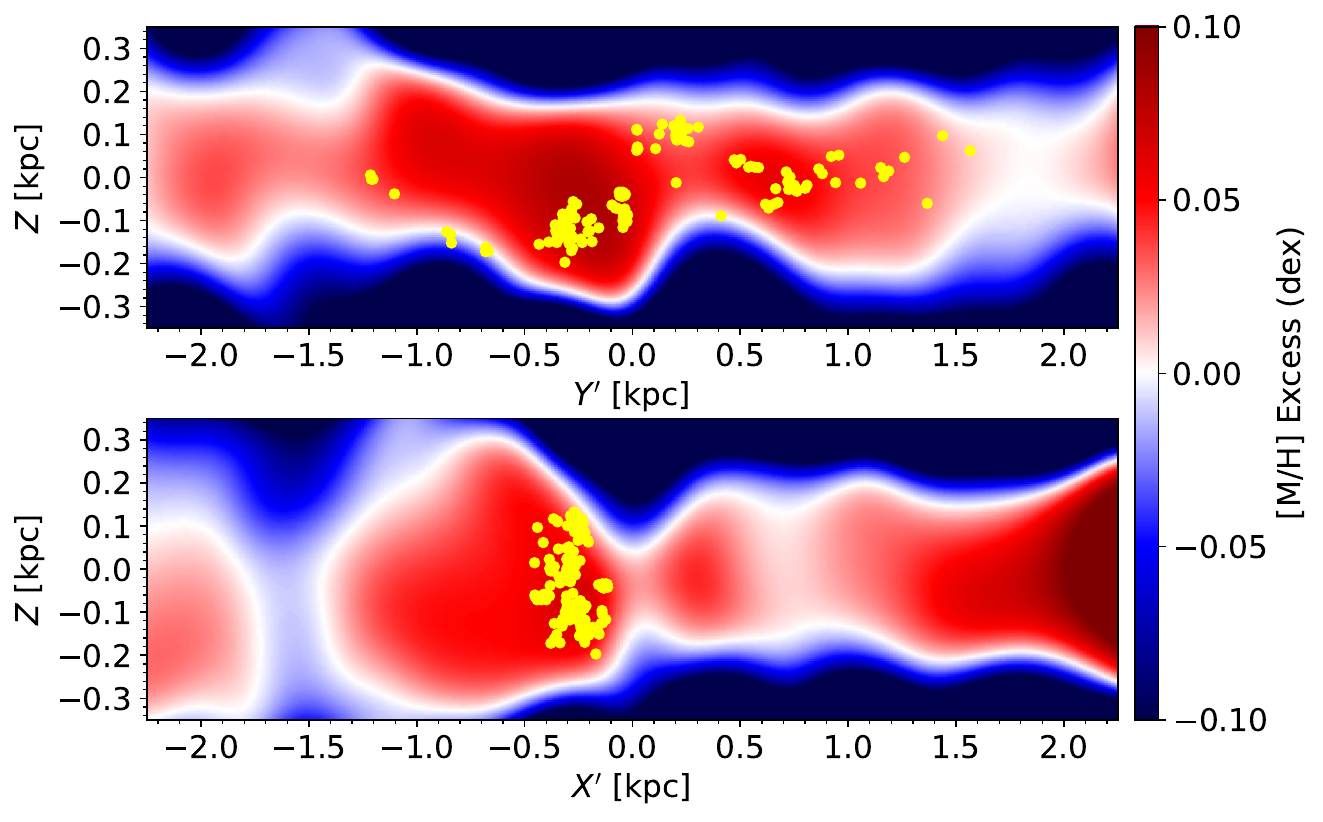}
\end{center}
\caption{{\it Top}: $Y^{\prime} - Z$ longitudinal cross-section of the 3D metallicity excess. The red band is a metal-rich stellar structure that we call the Extended Radcliffe Wave, since it undulates nearly in phase with the arrange of clouds complexes (yellow circles) from \citet[][]{Alves2020}. {\it Bottom}: $X^{\prime} - Z$ transversal cross-section of the 3D metallicity excess. The red oval-shaped structure centred at $X^{\prime}\approx-0.7\kpc$ is a transversal view of the Local Arm, while the Radcliffe Wave (yellow circles) is enclosed at the inner edge of the Local Arm. }
\label{fig:RW}
\end{figure*}

In Figure \ref{fig:RW} we show these longitudinal and transversal edge-on views and compare with the position of the Radcliffe Wave (array of yellow points). Note that in Figure \ref{fig:RW} (in the same way as in Figure \ref{fig:Arms_yz}) the [M/H] excess is not being projected, but it is rather a cross-section view of the volumetric distribution of [M/H] excess (see the interactive Figures for a reference\footnote[2]{\url{https://luismartinez-astro.github.io/Martinez-Medina_2025/Gallery.html}}). As expected, the $Y^{\prime}-Z$ projection shows a vertically undulating distribution of metal-rich stars. Even more remarkable is that this stellar wave is nearly in phase with the Radcliffe Wave, and is actually the lower boundary of the metal-rich stellar distribution the one that roughly follows the undulatory pattern of the Radcliffe Wave. We note, however, that the vertical distribution of the observed metallicity excess is broader than the Radcliffe Wave. Specifically, it extends to both positive and negative vertical coordinates around $Y^{\prime}\sim -0.5 - 0\kpc$, whereas the Radcliffe Wave is systematically shifted towards negative $Z$ (as shown in the top panel of Figure \ref{fig:RW}). This is evidence that the Radcliffe Wave belongs to a larger structure traced by young, dynamically cold, and metal-rich stars, which we call the Extended Radcliffe Wave; it reaches vertical amplitudes of nearly 270$\pc$, extends for at least 4$\kpc$, and has a maximum vertical thickness of $\sim400\pc$, containing entirely the original Radcliffe Wave.

Moreover, in the transversal cut of the 3D [M/H] excess (bottom panel of Figure \ref{fig:RW}) we also include the location of the Radcliffe Wave projected on the $X^{\prime}-Z$ plane. Note that these clouds complexes are enclosed within a prominent metal-rich structure centred at $X^{\prime}\approx-0.7\kpc$; which is a transversal cut of the Local Arm (as can be seen from Figure \ref{fig:MH_Excess_xy_rotated}), i.e., this is a confirmation that the Radcliffe Wave (or at least the part with $Y^{\prime}>0$) is contained by the Local Arm, as traced by stellar metallicity. Furthermore, the $X^{\prime}-Z$ plane shows that the Extended Radcliffe Wave is at the thin inner edge of the Local Arm, closer to the Sun's location.

Here it is worth mentioning that in the 2D map of Figure \ref{fig:MH_Excess_xy_rotated} the Radcliffe Wave appears to lie along the edge of the Local Arm rather than inside, and the explanation of this apparent discrepancy is the following: For the 2D projection the [M/H] excess at a given ($X,Y$) position is computed by integrating along the $Z$ axis, which is different to slicing through the 3D distribution of [M/H] excess. In particular, the bottom panel of Figure \ref{fig:RW} shows that the Local Arm as traced by metallicity excess is vertically thinner at its edges than at its centre, hence, when the 2D projection of Figure \ref{fig:MH_Excess_xy_rotated} is performed, the positive metallicity excess in these thin edges is dominated by the thicker negative excess above and below the edges of the arm, and the net excess at these positions will result negative in Figure \ref{fig:MH_Excess_xy_rotated}. Hence, the differences in Local Arm widths between Figures \ref{fig:MH_Excess_xy_rotated} and \ref{fig:RW} are a projection effect that hides the vertically thinner edges of the arm.

The finding of a metal-rich stellar structure undulating in phase with the Radcliffe Wave is a proof of concept for the method of extending, to the vertical direction, the strong correlation between positive [M/H] excess in the disc and spiral arms location. Moreover, the perturbation that gave origin to the Radcliffe Wave presumably affected a larger volume of the Galactic disc, not only in gas but also in its stellar component. This is in agreement with the results from stellar kinematics obtained by \citet{Thulasidharan:2022}.

\section{Conclusions}
\label{sec:conclusions} 

The morphology of the spiral arms can be better understood by using a variety of tracers. Nonetheless, although their projected location on the plane has been extensible studied, a mapping of the spiral arms in the vertical direction is a challenging task that has received little attention. 

As a relevant tracer we propose the three-dimensional chemical footprint of the spiral arms, specifically the use of the metallicity distribution of young and dynamically cold stars, as they constitute the backbone of the spiral arms. We decided to use a stellar sample where the metallicity excess has been proved to correlate with the location of the spiral arms and, based on this strong correlation, interpret the vertical metallicity distribution as a key diagnostic of the spiral arms in the vertical direction.

In this context, Figure \ref{fig:Elevation_Hillshade} presents a topographic map of the 3D morphology of the chemical signature of the arm segments within 4$\kpc$ from the Sun. This map shows spiral arms with a changing vertical thickness, a width that decreases away from the plane, and high asymmetry across the $Z=0$ plane. Additional details of the vertical structure of the arms are shown in Figure \ref{fig:Arms_yz}, which presents vertical cuts of the 3D metallicity excess along the Local, Sagittarius-Carina, Scutum, and Perseus arm segments. Here the asymmetry across the mid-plane is evident, especially for Perseus and Sagittarius-Carina, while Scutum is fairly symmetric. The observed vertical asymmetry in Perseus is presumably due to the Galactic warp, while the vertical fluctuations of the Sagittarius-Carina arm are less clear. It is possible that the observed variations (and clumpiness) are caused by interstellar extinction.

Our proposed method of tracing the three-dimensional metallicity distribution of stellar structures in the disc can be validated by applying it to vertical structures already well characterized by independent methods and tracers. One of these structures of interest, in the vicinity of the Sun, is the Radcliffe Wave \citep{Alves2020}. In the region where this structure is located, following the same procedure adopted for the spiral arms, we obtained a vertical cut of the 3D metallicity excess across a line defined by the position of the Radcliffe Wave in the mid-plane. Indeed, the $Y^{\prime}-Z$ edge-on view (top panel of Figure \ref{fig:RW}) reveals a metal-rich stellar structure that undulates vertically, roughly following the Radcliffe Wave. This structure, which we call the Extended Radcliffe Wave, extends for at least 4$\kpc$, reaches vertical amplitudes of nearly 270$\pc$, and has a maximum vertical thickness of $\sim400\pc$, containing entirely the original Radcliffe Wave.

Another important result is that in the transversal edge-on view (Figure \ref{fig:RW}) we confirm that the Extended Radcliffe Wave is at the inner edge of the Local Arm. This leaves open the question of what is the origin of this vertical perturbation, as it is just one part of the vertically undulating Local Arm. However, being located at the inner edge, our findings might be consistent with the scenario proposed by \cite{Swiggum:2022}, according to which the Radcliffe Wave is the gas reservoir of the Local (Orion) arm.

The finding of a metal-rich stellar counterpart of the Radcliffe Wave can be seen as a proof of concept, showing that mapping the three-dimensional metallicity distribution can reveal important clues about the structures, dynamics, and chemical enrichment in the Galactic disc.

\section*{Acknowledgements}

We thank the anonymous referee for their insightful suggestions, which contributed to improve this work. This work has made use of data from the European Space Agency (ESA) mission {\it Gaia} (\url{https://www.cosmos.esa.int/gaia}), processed by the {\it Gaia}
Data Processing and Analysis Consortium (DPAC, \url{https://www.cosmos.esa.int/web/gaia/dpac/consortium}). Funding for the DPAC
has been provided by national institutions, in particular the institutions
participating in the {\it Gaia} Multilateral Agreement.
LMM and EMH acknowledge support from DGAPA-PAPIIT IN108924 grant. EMH acknowledges support by the Tsinghua Shui Mu Scholarship. EP is supported in part by the Italian Space Agency (ASI) through contract 2018-24-HH.0 and its addendum 2018-24-HH.1-2022 to the National Institute for Astrophysics (INAF).
\section*{Data availability}

The data that support the findings of this study are available from the corresponding author, upon reasonable request.

\bibliographystyle{mnras}
\bibliography{ref}

\begin{thebibliography}{}
\makeatletter
\relax
\def\mn@urlcharsother{\let\do\@makeother \do\$\do\&\do\#\do\^\do\_\do\%\do\~}
\def\mn@doi{\begingroup\mn@urlcharsother \@ifnextchar [ {\mn@doi@} {\mn@doi@[]}}
\def\mn@doi@[#1]#2{\def\@tempa{#1}\ifx\@tempa\@empty \href {http://dx.doi.org/#2} {doi:#2}\else \href {http://dx.doi.org/#2} {#1}\fi \endgroup}
\def\mn@eprint#1#2{\mn@eprint@#1:#2::\@nil}
\def\mn@eprint@arXiv#1{\href {http://arxiv.org/abs/#1} {{\tt arXiv:#1}}}
\def\mn@eprint@dblp#1{\href {http://dblp.uni-trier.de/rec/bibtex/#1.xml} {dblp:#1}}
\def\mn@eprint@#1:#2:#3:#4\@nil{\def\@tempa {#1}\def\@tempb {#2}\def\@tempc {#3}\ifx \@tempc \@empty \let \@tempc \@tempb \let \@tempb \@tempa \fi \ifx \@tempb \@empty \def\@tempb {arXiv}\fi \@ifundefined {mn@eprint@\@tempb}{\@tempb:\@tempc}{\expandafter \expandafter \csname mn@eprint@\@tempb\endcsname \expandafter{\@tempc}}}

\bibitem[\protect\citeauthoryear{{Alves} et~al.,}{{Alves} et~al.}{2020}]{Alves2020}
{Alves} J.,  et~al., 2020, \mn@doi [\nat] {10.1038/s41586-019-1874-z}, \href {https://ui.adsabs.harvard.edu/abs/2020Natur.578..237A} {578, 237}

\bibitem[\protect\citeauthoryear{{Antoja} et~al.,}{{Antoja} et~al.}{2018}]{Antoja2018}
{Antoja} T.,  et~al., 2018, \mn@doi [\nat] {10.1038/s41586-018-0510-7}, \href {https://ui.adsabs.harvard.edu/abs/2018Natur.561..360A} {561, 360}

\bibitem[\protect\citeauthoryear{{Barbillon}, {Recio-Blanco}, {Poggio}, {Palicio}, {Spitoni}, {de Laverny}  \& {Cescutti}}{{Barbillon} et~al.}{2025}]{Barbillon2025}
{Barbillon} M.,  {Recio-Blanco} A.,  {Poggio} E.,  {Palicio} P.~A.,  {Spitoni} E.,  {de Laverny} P.,   {Cescutti} G.,  2025, \mn@doi [\aap] {10.1051/0004-6361/202450868}, \href {https://ui.adsabs.harvard.edu/abs/2025A&A...693A...3B} {693, A3}

\bibitem[\protect\citeauthoryear{{Barnes}, {Barnes}, {Hernandez}, {Lopez}  \& {Muller}}{{Barnes} et~al.}{2025}]{Barnes2025}
{Barnes} P.~J.,  {Barnes} D. G.~H.,  {Hernandez} A.~K.,  {Lopez} S.,   {Muller} E.,  2025, \mn@doi [arXiv e-prints] {10.48550/arXiv.2503.04887}, \href {https://ui.adsabs.harvard.edu/abs/2025arXiv250304887B} {p. arXiv:2503.04887}

\bibitem[\protect\citeauthoryear{{Belokurov}, {Erkal}, {Evans}, {Koposov}  \& {Deason}}{{Belokurov} et~al.}{2018}]{Belokurov2018}
{Belokurov} V.,  {Erkal} D.,  {Evans} N.~W.,  {Koposov} S.~E.,   {Deason} A.~J.,  2018, \mn@doi [\mnras] {10.1093/mnras/sty982}, \href {https://ui.adsabs.harvard.edu/abs/2018MNRAS.478..611B} {478, 611}

\bibitem[\protect\citeauthoryear{{Bennett} \& {Bovy}}{{Bennett} \& {Bovy}}{2019}]{Bennett2019}
{Bennett} M.,  {Bovy} J.,  2019, \mn@doi [\mnras] {10.1093/mnras/sty2813}, \href {https://ui.adsabs.harvard.edu/abs/2019MNRAS.482.1417B} {482, 1417}

\bibitem[\protect\citeauthoryear{{Cabrera-Gadea}, {Mateu}, {Ramos}, {Romero-G{\'o}mez}, {Antoja}  \& {Aguilar}}{{Cabrera-Gadea} et~al.}{2024}]{Cabrera-Gadea2024}
{Cabrera-Gadea} M.,  {Mateu} C.,  {Ramos} P.,  {Romero-G{\'o}mez} M.,  {Antoja} T.,   {Aguilar} L.,  2024, \mn@doi [\mnras] {10.1093/mnras/stae308}, \href {https://ui.adsabs.harvard.edu/abs/2024MNRAS.528.4409C} {528, 4409}

\bibitem[\protect\citeauthoryear{{Carr}, {Johnston}, {Laporte}  \& {Ness}}{{Carr} et~al.}{2022}]{Carr2022}
{Carr} C.,  {Johnston} K.~V.,  {Laporte} C. F.~P.,   {Ness} M.~K.,  2022, \mn@doi [\mnras] {10.1093/mnras/stac2403}, \href {https://ui.adsabs.harvard.edu/abs/2022MNRAS.516.5067C} {516, 5067}

\bibitem[\protect\citeauthoryear{{Castro-Ginard} et~al.,}{{Castro-Ginard} et~al.}{2021}]{Castro-Ginard2021}
{Castro-Ginard} A.,  et~al., 2021, \mn@doi [\aap] {10.1051/0004-6361/202039751}, \href {https://ui.adsabs.harvard.edu/abs/2021A&A...652A.162C} {652, A162}

\bibitem[\protect\citeauthoryear{{Debattista}, {Khachaturyants}, {Amarante}, {Carr}, {Beraldo e Silva}  \& {Laporte}}{{Debattista} et~al.}{2025}]{Debattista2025}
{Debattista} V.~P.,  {Khachaturyants} T.,  {Amarante} J. A.~S.,  {Carr} C.,  {Beraldo e Silva} L.,   {Laporte} C. F.~P.,  2025, \mn@doi [\mnras] {10.1093/mnras/staf035}, \href {https://ui.adsabs.harvard.edu/abs/2025MNRAS.537.1620D} {537, 1620}

\bibitem[\protect\citeauthoryear{{Dehnen}, {Semczuk}  \& {Sch{\"o}nrich}}{{Dehnen} et~al.}{2023}]{Dehnen2023}
{Dehnen} W.,  {Semczuk} M.,   {Sch{\"o}nrich} R.,  2023, \mn@doi [\mnras] {10.1093/mnras/stad1502}, \href {https://ui.adsabs.harvard.edu/abs/2023MNRAS.523.1556D} {523, 1556}

\bibitem[\protect\citeauthoryear{{Di Matteo}, {Haywood}, {Combes}, {Semelin}  \& {Snaith}}{{Di Matteo} et~al.}{2013}]{DiMatteo2013}
{Di Matteo} P.,  {Haywood} M.,  {Combes} F.,  {Semelin} B.,   {Snaith} O.~N.,  2013, \mn@doi [\aap] {10.1051/0004-6361/201220539}, \href {https://ui.adsabs.harvard.edu/abs/2013A&A...553A.102D} {553, A102}

\bibitem[\protect\citeauthoryear{{Eilers}, {Hogg}, {Rix}, {Frankel}, {Hunt}, {Fouvry}  \& {Buck}}{{Eilers} et~al.}{2020}]{Eilers2020}
{Eilers} A.-C.,  {Hogg} D.~W.,  {Rix} H.-W.,  {Frankel} N.,  {Hunt} J. A.~S.,  {Fouvry} J.-B.,   {Buck} T.,  2020, \mn@doi [\apj] {10.3847/1538-4357/abac0b}, \href {https://ui.adsabs.harvard.edu/abs/2020ApJ...900..186E} {900, 186}

\bibitem[\protect\citeauthoryear{{Filion}, {McClure}, {Weinberg}, {D'Onghia}  \& {Daniel}}{{Filion} et~al.}{2023}]{Filion2023}
{Filion} C.,  {McClure} R.~L.,  {Weinberg} M.~D.,  {D'Onghia} E.,   {Daniel} K.~J.,  2023, \mn@doi [\mnras] {10.1093/mnras/stad1832}, \href {https://ui.adsabs.harvard.edu/abs/2023MNRAS.524..276F} {524, 276}

\bibitem[\protect\citeauthoryear{{GRAVITY Collaboration} et~al.,}{{GRAVITY Collaboration} et~al.}{2021}]{GRAVITY2021}
{GRAVITY Collaboration} et~al., 2021, \mn@doi [\aap] {10.1051/0004-6361/202140981}, \href {https://ui.adsabs.harvard.edu/abs/2021A&A...654A..22G} {654, A22}

\bibitem[\protect\citeauthoryear{{Gaia Collaboration} et~al.,}{{Gaia Collaboration} et~al.}{2016}]{GaiaCollaboration:2016}
{Gaia Collaboration} et~al., 2016, \mn@doi [\aap] {10.1051/0004-6361/201629512}, \href {https://ui.adsabs.harvard.edu/abs/2016A&A...595A...2G} {595, A2}

\bibitem[\protect\citeauthoryear{{Gaia Collaboration} et~al.,}{{Gaia Collaboration} et~al.}{2018}]{GaiaCollaboration:2018}
{Gaia Collaboration} et~al., 2018, \mn@doi [\aap] {10.1051/0004-6361/201833051}, \href {https://ui.adsabs.harvard.edu/abs/2018A&A...616A...1G} {616, A1}

\bibitem[\protect\citeauthoryear{{Gaia Collaboration} et~al.,}{{Gaia Collaboration} et~al.}{2021}]{GaiaCollaboration:2021}
{Gaia Collaboration} et~al., 2021, \mn@doi [\aap] {10.1051/0004-6361/202039657}, \href {https://ui.adsabs.harvard.edu/abs/2021A&A...649A...1G} {649, A1}

\bibitem[\protect\citeauthoryear{{Gaia Collaboration} et~al.,}{{Gaia Collaboration} et~al.}{2023a}]{GaiaCollaboration:2023}
{Gaia Collaboration} et~al., 2023a, \mn@doi [\aap] {10.1051/0004-6361/202243940}, \href {https://ui.adsabs.harvard.edu/abs/2023A&A...674A...1G} {674, A1}

\bibitem[\protect\citeauthoryear{{Gaia Collaboration} et~al.,}{{Gaia Collaboration} et~al.}{2023b}]{Recio-Blanco2023}
{Gaia Collaboration} et~al., 2023b, \mn@doi [\aap] {10.1051/0004-6361/202243511}, \href {https://ui.adsabs.harvard.edu/abs/2023A&A...674A..38G} {674, A38}

\bibitem[\protect\citeauthoryear{{Graf}, {Wetzel}, {Bellardini}  \& {Bailin}}{{Graf} et~al.}{2025}]{Graf2025}
{Graf} R.~L.,  {Wetzel} A.,  {Bellardini} M.~A.,   {Bailin} J.,  2025, \mn@doi [\apj] {10.3847/1538-4357/adacd7}, \href {https://ui.adsabs.harvard.edu/abs/2025ApJ...981...47G} {981, 47}

\bibitem[\protect\citeauthoryear{{Grand} et~al.,}{{Grand} et~al.}{2016}]{Grand:2016}
{Grand} R. J.~J.,  et~al., 2016, \mn@doi [\mnras] {10.1093/mnrasl/slw086}, \href {https://ui.adsabs.harvard.edu/abs/2016MNRAS.460L..94G} {460, L94}

\bibitem[\protect\citeauthoryear{{Hackshaw}, {Hawkins}, {Filion}, {Horta}, {Laporte}, {Carr}  \& {Price-Whelan}}{{Hackshaw} et~al.}{2024}]{Hackshaw2024}
{Hackshaw} Z.,  {Hawkins} K.,  {Filion} C.,  {Horta} D.,  {Laporte} C. F.~P.,  {Carr} C.,   {Price-Whelan} A.~M.,  2024, \mn@doi [\apj] {10.3847/1538-4357/ad900e}, \href {https://ui.adsabs.harvard.edu/abs/2024ApJ...977..143H} {977, 143}

\bibitem[\protect\citeauthoryear{{Hawkins}}{{Hawkins}}{2023}]{Hawkins2023}
{Hawkins} K.,  2023, \mn@doi [\mnras] {10.1093/mnras/stad1244}, \href {https://ui.adsabs.harvard.edu/abs/2023MNRAS.525.3318H} {525, 3318}

\bibitem[\protect\citeauthoryear{{Hunt} \& {Vasiliev}}{{Hunt} \& {Vasiliev}}{2025}]{Hunt_Vasiliev2025}
{Hunt} J. A.~S.,  {Vasiliev} E.,  2025, \mn@doi [\nar] {10.1016/j.newar.2024.101721}, \href {https://ui.adsabs.harvard.edu/abs/2025NewAR.10001721H} {100, 101721}

\bibitem[\protect\citeauthoryear{{Hunt}, {Price-Whelan}, {Johnston}, {McClure}, {Filion}, {Cassese}  \& {Horta}}{{Hunt} et~al.}{2024}]{Hunt2024}
{Hunt} J. A.~S.,  {Price-Whelan} A.~M.,  {Johnston} K.~V.,  {McClure} R.~L.,  {Filion} C.,  {Cassese} B.,   {Horta} D.,  2024, \mn@doi [\mnras] {10.1093/mnras/stad3918}, \href {https://ui.adsabs.harvard.edu/abs/2024MNRAS.52711393H} {527, 11393}

\bibitem[\protect\citeauthoryear{{Khoperskov} \& {Gerhard}}{{Khoperskov} \& {Gerhard}}{2022}]{Khoperskov2022}
{Khoperskov} S.,  {Gerhard} O.,  2022, \mn@doi [\aap] {10.1051/0004-6361/202141836}, \href {https://ui.adsabs.harvard.edu/abs/2022A&A...663A..38K} {663, A38}

\bibitem[\protect\citeauthoryear{{Khoperskov}, {Di Matteo}, {Haywood}  \& {Combes}}{{Khoperskov} et~al.}{2018}]{Khoperskov2018}
{Khoperskov} S.,  {Di Matteo} P.,  {Haywood} M.,   {Combes} F.,  2018, \mn@doi [\aap] {10.1051/0004-6361/201732521}, \href {https://ui.adsabs.harvard.edu/abs/2018A&A...611L...2K} {611, L2}

\bibitem[\protect\citeauthoryear{{Khoperskov}, {Sivkova}, {Saburova}, {Vasiliev}, {Shustov}, {Minchev}  \& {Walcher}}{{Khoperskov} et~al.}{2023}]{Khoperskov2023}
{Khoperskov} S.,  {Sivkova} E.,  {Saburova} A.,  {Vasiliev} E.,  {Shustov} B.,  {Minchev} I.,   {Walcher} C.~J.,  2023, \mn@doi [\aap] {10.1051/0004-6361/202142581}, \href {https://ui.adsabs.harvard.edu/abs/2023A&A...671A..56K} {671, A56}

\bibitem[\protect\citeauthoryear{{Laporte}, {Minchev}, {Johnston}  \& {G{\'o}mez}}{{Laporte} et~al.}{2019}]{Laporte2019}
{Laporte} C. F.~P.,  {Minchev} I.,  {Johnston} K.~V.,   {G{\'o}mez} F.~A.,  2019, \mn@doi [\mnras] {10.1093/mnras/stz583}, \href {https://ui.adsabs.harvard.edu/abs/2019MNRAS.485.3134L} {485, 3134}

\bibitem[\protect\citeauthoryear{{Lemasle} et~al.,}{{Lemasle} et~al.}{2022}]{Lemasle2022}
{Lemasle} B.,  et~al., 2022, \mn@doi [\aap] {10.1051/0004-6361/202243273}, \href {https://ui.adsabs.harvard.edu/abs/2022A&A...668A..40L} {668, A40}

\bibitem[\protect\citeauthoryear{{Lian}, {Zasowski}, {Chen}, {Imig}, {Wang}, {Boardman}  \& {Liu}}{{Lian} et~al.}{2024}]{Lian2024}
{Lian} J.,  {Zasowski} G.,  {Chen} B.,  {Imig} J.,  {Wang} T.,  {Boardman} N.,   {Liu} X.,  2024, \mn@doi [Nature Astronomy] {10.1038/s41550-024-02315-7}, \href {https://ui.adsabs.harvard.edu/abs/2024NatAs...8.1302L} {8, 1302}

\bibitem[\protect\citeauthoryear{{Martinez-Medina}, {P{\'e}rez-Villegas}  \& {Peimbert}}{{Martinez-Medina} et~al.}{2022}]{Martinez-Medina2022}
{Martinez-Medina} L.,  {P{\'e}rez-Villegas} A.,   {Peimbert} A.,  2022, \mn@doi [\mnras] {10.1093/mnras/stac642}, \href {https://ui.adsabs.harvard.edu/abs/2022MNRAS.512.1574M} {512, 1574}

\bibitem[\protect\citeauthoryear{{Palicio}, {Recio-Blanco}, {Poggio}, {Antoja}, {McMillan}  \& {Spitoni}}{{Palicio} et~al.}{2023}]{Palicio2023}
{Palicio} P.~A.,  {Recio-Blanco} A.,  {Poggio} E.,  {Antoja} T.,  {McMillan} P.~J.,   {Spitoni} E.,  2023, \mn@doi [\aap] {10.1051/0004-6361/202245026}, \href {https://ui.adsabs.harvard.edu/abs/2023A&A...670L...7P} {670, L7}

\bibitem[\protect\citeauthoryear{{Poggio, E.} et~al.,}{{Poggio, E.} et~al.}{2025}]{Poggio2024}
{Poggio, E.} et~al., 2025, \mn@doi [A&A] {10.1051/0004-6361/202451668}, 699, A199

\bibitem[\protect\citeauthoryear{{Poggio} et~al.,}{{Poggio} et~al.}{2021}]{Poggio2021}
{Poggio} E.,  et~al., 2021, \mn@doi [\aap] {10.1051/0004-6361/202140687}, \href {https://ui.adsabs.harvard.edu/abs/2021A&A...651A.104P} {651, A104}

\bibitem[\protect\citeauthoryear{{Poggio} et~al.,}{{Poggio} et~al.}{2022}]{Poggio2022}
{Poggio} E.,  et~al., 2022, \mn@doi [\aap] {10.1051/0004-6361/202244361}, \href {https://ui.adsabs.harvard.edu/abs/2022A&A...666L...4P} {666, L4}

\bibitem[\protect\citeauthoryear{{Reid} \& {Brunthaler}}{{Reid} \& {Brunthaler}}{2020}]{Reid2020}
{Reid} M.~J.,  {Brunthaler} A.,  2020, \mn@doi [\apj] {10.3847/1538-4357/ab76cd}, \href {https://ui.adsabs.harvard.edu/abs/2020ApJ...892...39R} {892, 39}

\bibitem[\protect\citeauthoryear{{Spitoni}, {Cescutti}, {Minchev}, {Matteucci}, {Silva Aguirre}, {Martig}, {Bono}  \& {Chiappini}}{{Spitoni} et~al.}{2019}]{Spitoni2019}
{Spitoni} E.,  {Cescutti} G.,  {Minchev} I.,  {Matteucci} F.,  {Silva Aguirre} V.,  {Martig} M.,  {Bono} G.,   {Chiappini} C.,  2019, \mn@doi [\aap] {10.1051/0004-6361/201834665}, \href {https://ui.adsabs.harvard.edu/abs/2019A&A...628A..38S} {628, A38}

\bibitem[\protect\citeauthoryear{{Spitoni} et~al.,}{{Spitoni} et~al.}{2023}]{Spitoni2023}
{Spitoni} E.,  et~al., 2023, \mn@doi [\aap] {10.1051/0004-6361/202347325}, \href {https://ui.adsabs.harvard.edu/abs/2023A&A...680A..85S} {680, A85}

\bibitem[\protect\citeauthoryear{{Swiggum} et~al.,}{{Swiggum} et~al.}{2022}]{Swiggum:2022}
{Swiggum} C.,  et~al., 2022, \mn@doi [\aap] {10.1051/0004-6361/202243761}, \href {https://ui.adsabs.harvard.edu/abs/2022A&A...664L..13S} {664, L13}

\bibitem[\protect\citeauthoryear{{Thulasidharan}, {D'Onghia}, {Poggio}, {Drimmel}, {Gallagher}, {Swiggum}, {Benjamin}  \& {Alves}}{{Thulasidharan} et~al.}{2022}]{Thulasidharan:2022}
{Thulasidharan} L.,  {D'Onghia} E.,  {Poggio} E.,  {Drimmel} R.,  {Gallagher} III J.~S.,  {Swiggum} C.,  {Benjamin} R.~A.,   {Alves} J.,  2022, \mn@doi [\aap] {10.1051/0004-6361/202142899}, \href {https://ui.adsabs.harvard.edu/abs/2022A&A...660L..12T} {660, L12}

\bibitem[\protect\citeauthoryear{{Vasiliev}, {Belokurov}  \& {Erkal}}{{Vasiliev} et~al.}{2021}]{Vasiliev2021}
{Vasiliev} E.,  {Belokurov} V.,   {Erkal} D.,  2021, \mn@doi [\mnras] {10.1093/mnras/staa3673}, \href {https://ui.adsabs.harvard.edu/abs/2021MNRAS.501.2279V} {501, 2279}

\makeatother
\end{thebibliography}

\appendix

\section{Effect of the smoothing scale-lengths in [M/H] excess}
\label{appendix:smoothing}

As described in Section \ref{sec:data}, the metallicity excess is defined as the difference between the local metallicity and the metallicity on a large scale, both quantities are averaged using a smoothing Gaussian kernel. The local metallicity is computed over a scale-length of $h_{\rm loc}=0.175\kpc$, while for the large scale we adopt the value $h_{\rm large}=7\times h_{\rm loc}$.

\begin{figure*}
\begin{center}
\includegraphics[width=\textwidth]{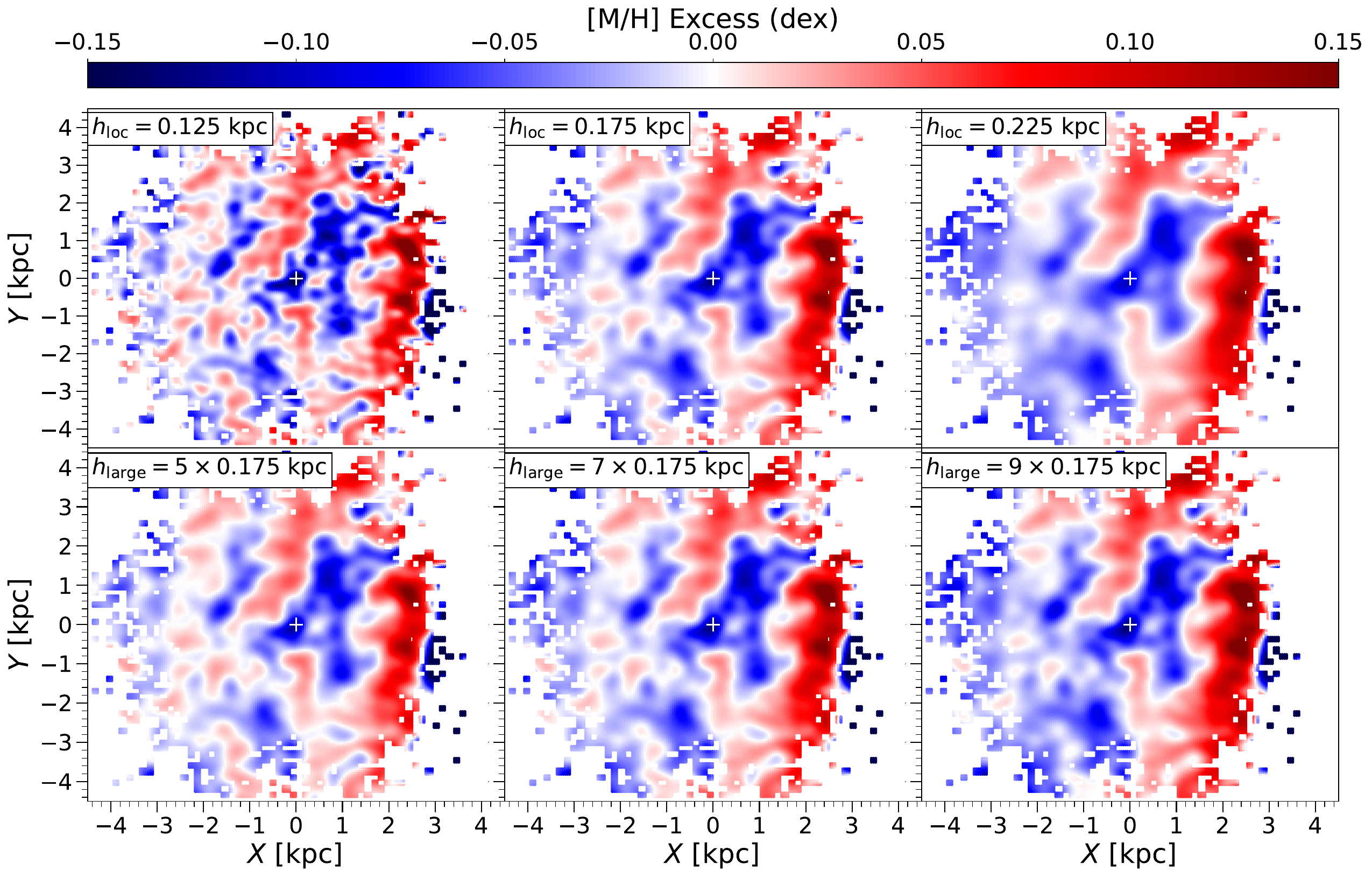}
\end{center}
\caption{{\it Top row}: [M/H] excess with three different values of the local scale-length, $h_{\rm loc}$, used to make the Gaussian smoothing. In all cases $h_{\rm large}=7\times h_{\rm loc}$. {\it Bottom row}: [M/H] excess with three different values of the large scale-length $h_{\rm large}$. In all cases $h_{\rm loc}=0.175\kpc$.}
\label{fig:smoothing_A}
\end{figure*}

In Figure \ref{fig:smoothing_A} we explore other values for the local scale-length (top row) and for the proportionality factor between the large and local scale-lengths (bottom row). In the top row note that smaller values of $h_{\rm loc}$ increase the level of noise and the structures appear fragmented compared to our adopted value (central panel). On the other hand, large values of $h_{\rm loc}$ oversmooth the small substructures since it no longer represents a local measurement of the metallicity. Note that for these variations of $h_{\rm loc}$ the large scale-length is always $h_{\rm large}=7\times h_{\rm loc}$.

In the bottom row of Figure \ref{fig:smoothing_A}, where we fix the value of $h_{\rm loc}$ and vary the proportional factor that relates $h_{\rm loc}$ and $h_{\rm large}$, we find a similar trend. For a factor of 5 the metallicity excess appears slightly more clumpy, mainly in the third Galactic quadrant ($X<0$ and $Y<0$); this is because at these values $h_{\rm large}$ is not large enough to provide a measurement of the mean metallicity on a large scale. On the other hand, for a value of 9 the small structures start to disappear and become oversmoothed. Nonetheless, for the cases explored here the larger structures that correspond to the arm segments analysed in this work are conserved, and our adopted values of $h_{\rm loc}=0.175\kpc$ and $h_{\rm large}=7\times h_{\rm loc}$ are a compromise to avoid noise and oversmoothed structures.

\section{Uncertainties in $\langle[{\rm M/H}]\rangle$}

\begin{figure*}
\begin{center}
\includegraphics[width=\textwidth]{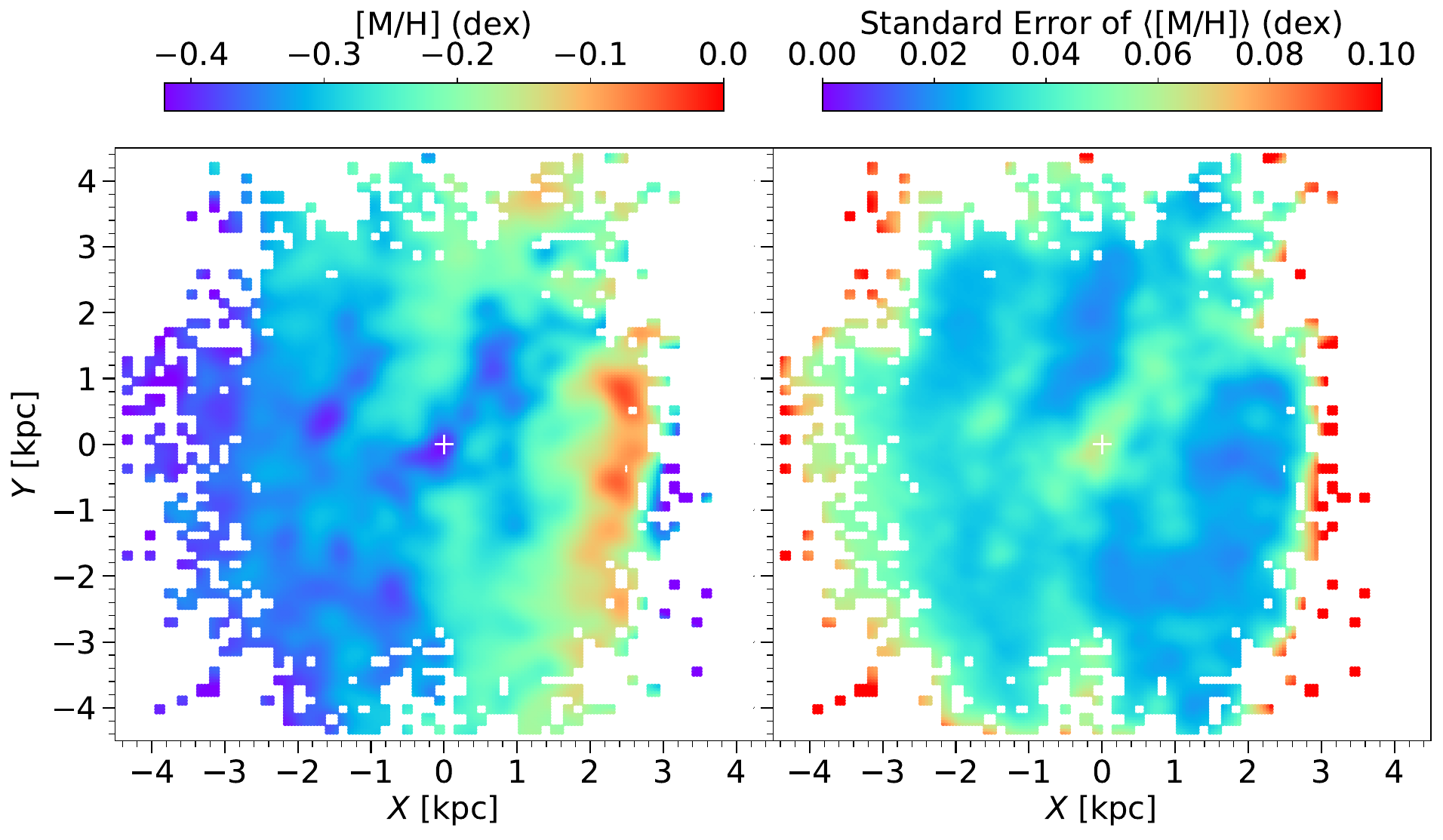}
\end{center}
\caption{{\it Left panel}: 2D mean metallicity distribution for our stellar sample showing the azimuthal inhomogeneities first noticed in \citetalias{Poggio2022}. {\it Right panel}: Standard error of the mean metallicity for our stellar sample, showing low uncertainties in the determination of $\langle[{\rm M/H}]\rangle$, particularly along the spiral arms.\label{lastpage}}
\label{fig:MHerr}
\end{figure*}

In this section we analyse the uncertainties and statistical significance of the substructures present in the metallicity excess map (see Figure \ref{fig:MH_Excess_xy}). To make this analysis, first note that the regions of positive metallicity excess are also visible in the metallicity distribution of the stellar sample (Figure 1 from \citetalias{Poggio2022}). In the left panel of Figure \ref{fig:MHerr} we reproduce such map of mean metallicity to show that the studied spiral arm segments stand out as azimuthal inhomogeneities. Hence, it is more practical and clearer to compute the uncertainties of the substructures in the map of mean [M/H].

In the right panel of Figure \ref{fig:MHerr} we show the standard error of the mean metallicity, which accounts for the variance around the mean values of [M/H] and the spatial density of stars. The first thing to notice is that in general the uncertainties in mean metallicity are small compared to the mean metallicity, this is because the region within 4$\kpc$ from the Sun is well sampled. The exception are the outskirts of our sample, which are less populated mainly due to the quality constraints applied to obtain this sample. The second result from the error map is that the regions of the spiral arms, being overdensities of stars, show good determinations of the mean metallicity, with errors less than $\sim 0.02$ dex. Hence, since the regions that correlate with the position of the spiral arms have mean metallicities of $\sim -0.2$ dex or larger, the relative error in mean metallicity for the metal-rich structures is around 10\%. These low relative errors, combined with the high-quality chemical measurements of individual stars in our sample, ensure good determinations of our metallicity maps and the structures in them.

\bsp

\end{document}